\documentclass[12pt,preprint]{aastex}

\usepackage{epsfig}
\shorttitle{Observability of Vortices}
\shortauthors{S.\ Wolf \& H.\ Klahr}

\begin{document}


\title{Large-scale Vortices in Protoplanetary Disks:\\
On the observability of possible early stages of planet formation}


\author{Sebastian Wolf}
\affil{California Institute of Technology, 1200 E California Blvd, Mail code 220-6, Pasadena, CA 91125, USA}
\email{swolf@ipac.caltech.edu}
%\and

\author{H. Klahr\altaffilmark{1}}
\affil{Universit\"at T\"ubingen, Institut f\"ur Astronomie und Astrophysik,
Abt. Computational Physics, Auf~der~Morgenstelle~10,
D-72076 T\"ubingen, Germany}
\email{hubert.klahr@uni-tuebingen.de}

\altaffiltext{1}{also at: UCO/Lick Observatory, University of California,
    Santa Cruz, CA 95060}

\begin{abstract}
We investigate the possibility of mapping large-scale anti-cyclonic vortices,
resulting from a global baroclinic instability,
as pre-cursors of planet formation in proto-planetary disks with the planned
Atacama Large Millimeter Array (ALMA). 
On the basis of three-dimensional radiative transfer simulations, images of a hydrodynamically
calculated disk are derived which provide the basis for the simulation of ALMA.
We find that ALMA will be able to trace the theoretically predicted
large-scale  anti-cyclonic vortex and will therefore allow testing of existing
models of this very early stage of planet formation in circumstellar disks.
\end{abstract}



%-------------------------------------------------------------------------------
\section{Introduction}\label{intro}

Klahr \& Bodenheimer~(2002) show that a global baroclinic instability represents
a source for turbulence leading to angular momentum transport in Keplerian
accretion disks with a radial gradient in entropy.
Their hydrodynamical simulations show that this baroclinic flow
is unstable and produces 
pressure waves, Rossby waves, and vortices in the $r-\phi$ plane of the disk.
Most interestingly, these hyper-dens anti-cyclonic vortices form out of little background noise and
become long-lasting features, which have been suggested to lead to the formation of planets (e.g.\ Adams \& Watkins 1995).
These pre-planetary matter concentrations have a surface density up to four times
higher then the ambient medium. They furthermore concentrate dust in their centres,
stressing their importance for the planetary formation process (e.g.\ Barge \& Sommeria 1995).

%-------------------------------------------------------------------------------
\section{Hydrodynamic and radiative transfer simulations}\label{model}

Using the code TRAMP (Klahr et al.~1999), we perform 2D hydrodynamic simulations 
of the evolution of the inner region (1--10\,AU) of a circumstellar disk with 
an initial mass of $\approx 2\times10^{-3}\,{\rm M_{\sun}}$.
The comparison with previous models of circumstellar disks around T~Tauri stars shows
(see, e.g., Wood et al.~2002, Cotera et al.~2001) that
this corresponds to a total disk mass on the order of $10^{-2}\,{\rm M_{\sun}}$.
As described by Klahr \& Bodenheimer~(2002) we treat the evolution of the 
vertically integrated density $\Sigma$ and internal energy $\sim \Sigma T_{\rm cent}$. The initial model 
had a slope in surface density of $R^{-0.75}$ and a temperature slope
of $R^{-1.0}$ which represents a constant pressure scale height of $H/r = 0.055$. 
The vertically integrated pressure is then given
by the ideal gas equation. Sources $Q^+$ for internal energy are the usual
$PdV$-work as well as the dissipation of possible shocks via
a simple von Neumann-Richtmyer viscosity (see Stone \& Norman 1992).
A sink for the internal energy is a black body cooling function 
$Q^-=a T_{\rm eff}^{4}$ mimicking the effect of optically thick radiation in the disk and cooling with 
the effective (surface) temperature $T_{\rm eff} = T_{\rm cent} (4/3 \tau)^{1/4}$
with the optical depth $\tau$ as a function of temperature and surface density. 
The simulation uses 128 cells in the azimuthal
direction and 128 logarithmically distributed cells in the radial direction.
The initial distribution was disturbed by a $0.1$ percent perturbation in density
and left to evolve freely afterwards under the influence of the global
radial entropy gradient. After $10^4$ yrs a huge anti-cyclonic hyper-dens vortex
has formed (see Klahr \& Bodenheimer ~2002). Afterwards the 3D data was
re-established by assuming a local vertical pressure equilibrium.

Based on the density and temperature distributions
we performed radiative transfer simulations using the three-dimensional
continuum radiative transfer code MC3D (Wolf~2002; see also Wolf et al.~1999,
Wolf \& Henning~2000). The goal of these simulations was to obtain images
at frequencies of 345\,GHz and 900\,GHz.
While the first frequency was chosen in order to allow comparison with
former investigations focused mainly on this 
frequency, 
the second frequency marks the planned upper limit of
the frequency range to be covered by 
ALMA\footnote{See {\tt http://www.arcetri.astro.it/science/ALMA/}
for a compilation of ALMA related documents.} 
and is of special importance for
the observation of small spatial structures. The resulting images provide
the basis for simulations of the continuum observations with ALMA
discussed in \S\ref{simmial}.

Since we consider continuum observations at submillimeter wavelengths, 
the circumstellar dust is the dominant source of emission.
The dust grains are assumed spherical, consisting of a mixture of 62.5~\% astronomical silicate
and 37.5~\% graphite (optical data from Weingartner \& Draine~2001\footnote{See also
{\tt http://www.astro.princeton.edu/$\tilde{ }$ draine}.}).
For the graphite we adopt the usual ``$\frac{1}{3}-\frac{2}{3}$'' approximation 
(Draine \& Malhotra~1993).
%:
%$Q_{\rm ext} = [Q_{\rm ext}(\epsilon_{\parallel}) + 2 Q_{\rm ext}(\epsilon_{\perp})]/3$,
%where $\epsilon_{\parallel}$ and $\epsilon_{\perp}$ are the components of the graphite
%dielectric tensor for the electric field parallel and perpendicular to the crystallographic $c$-axis,
%respectively ($Q_{\rm ext}$ is the extinction efficiency factor).
The size distribution $n(a)$ of the grains follows a power law, $n(a) \propto a^{-3.5}$, with
grain radii in the range $0.005\,\mu{\rm m} \le a \le 1\,\mu{\rm m}$.
The gas-to-dust mass ratio amounts to 100:1 in our simulations.

%-------------------------------------------------------------------------------
\section{Simulations of observations with ALMA}\label{simmial}

For simulation of the observations with ALMA we use the simulator
software published by Pety et al.~(2001). Since the goal of the observation
is to find small-scale structures in circumstellar disks, we consider
the ALMA configuration with the longest baselines (max.\ baseline: $\approx$13\,km)
in which the 64 12m-antennas are distributed in rings. 
In accordance with the suggestions given by Pety et al.~(2001), the following main types 
of errors are introduced in order to make the simulations realistic:
(a) frequency-dependent receiver- and system temperatures as given by Guilloteau~(2002),
(b) random pointing error during the observation with a maximum value of of 0.6'' in each direction,
and
(c) $30^{\rm o}$ phase noise. Further error sources, such as amplitude errors and 
``anomalous'' refraction (due to the variation of the refractive index of the wet air along
the line of sight) are considered as well.
The observations are simulated for continuous observations centered on the meridian transit
of the object. The object passes the meridian in zenith where an opacity of 0.15 is assumed.
The bandwidth amounts to 8\,GHz.

\begin{figure}[ht]
  \epsscale{0.6}
  \plotone{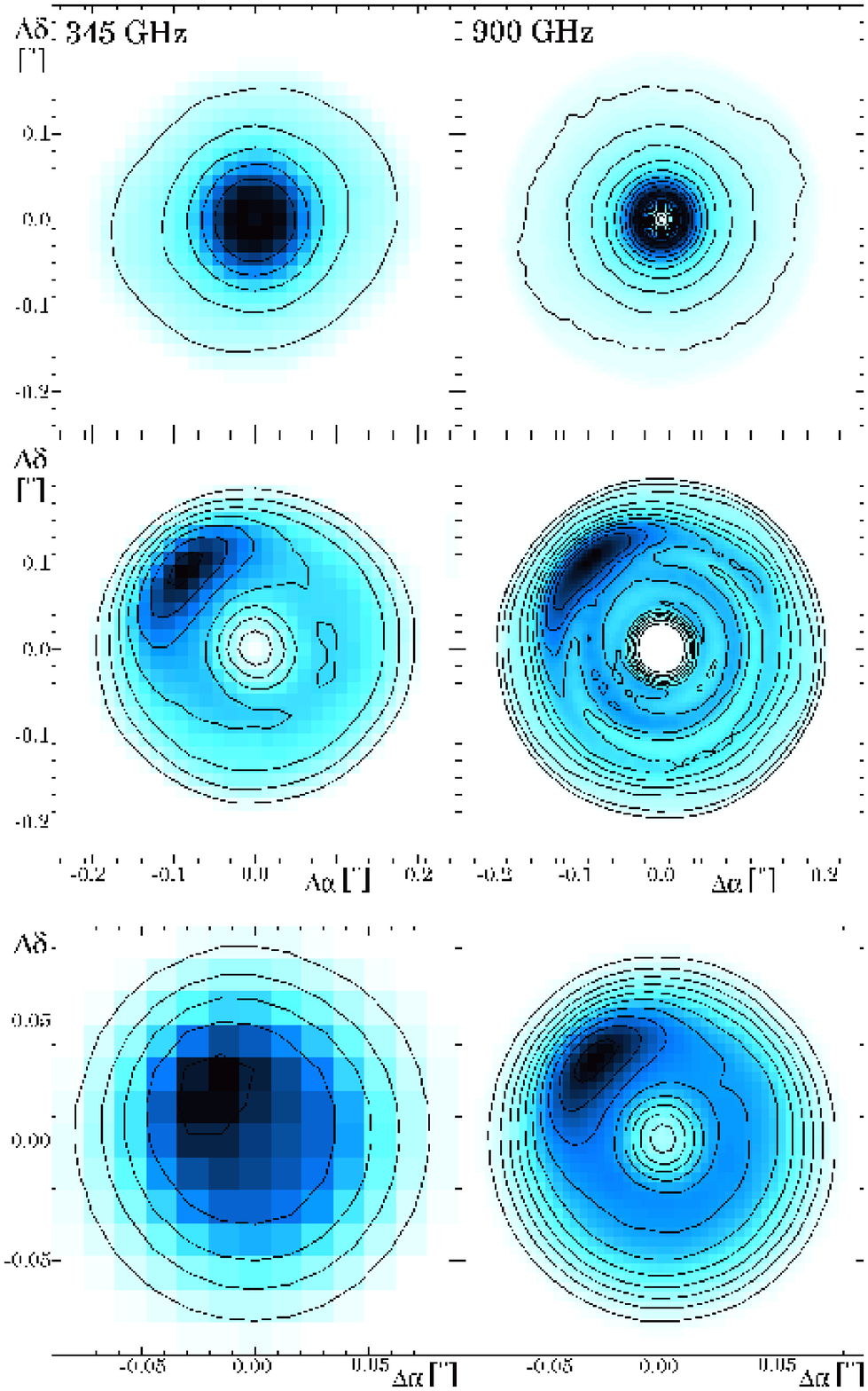}
  \caption{Reconstructed images of the disk seen face-on
        at $\nu$=345\,GHz (left column) and 900\,GHz (right column).
        {\sl Upper row:} Undisturbed disk, distance 50\,pc.
        {\sl Middle/lower row:} Evolved disk with baroclinic instability, 
        distance 50\,pc/140\,pc. 
        The contour lines mark steps of 0.5\,mJy/beam and 1.5\,mJy/beam
        at $\nu$=345\,GHz and $\nu$=900\,GHz, respectively.
        Total integration time: 2\,h.}
  \label{pic56-4}
\end{figure}

\begin{figure*}[ht]
  \epsscale{1.0}
  \plotone{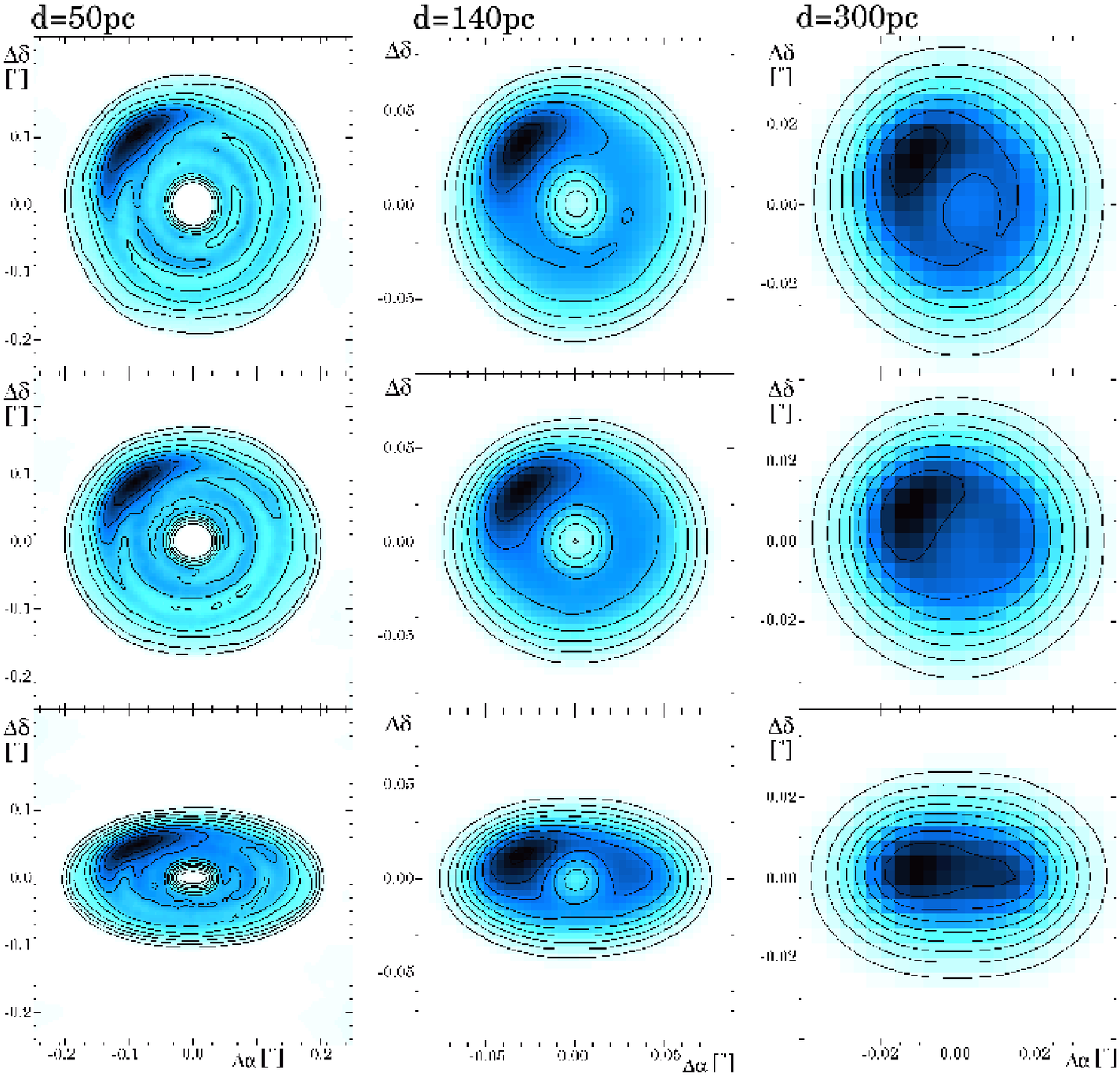}
  \caption{Reconstructed images of the evolved disk seen under inclinations of
        {\sl i}=$0^{\rm o}$ (face-on; upper row),
        $30^{\rm o}$ (middle row), and
        $60^{\rm o}$ (lower row). Distances of 50\,pc, 140\,pc, and 300\,pc are considered.
        $\nu$=900\,GHz;
        total integration time: 2\,h.
        The contour lines mark steps of 2.0\,mJy/beam${^{(1)}}$ and 1.5\,mJy/beam${^{(2)}}$
        in the case of {\sl d}=50\,pc${^{(1)}}$, 140\,pc${^{(1)}}$ and 300\,pc${^{(2)}}$, respectively.}
  \label{pic1-all3}
\end{figure*}

Fig.~\ref{pic56-4} shows simulated images of the undisturbed disk in comparison
to the disk with the evolved baroclinic instability at 345\,GHz and 900\,GHz
for object distances of 50\,pc and 140\,pc. The hyper-dens vortex is clearly visible
even at the lower frequency. However, in the case of an assumed distance of 140\,pc, 
the vortex causes a slight asymmetry in the reconstructed image only and could be hardly
identified. Since several well-studied nearby star-forming regions which can be observed
from the prospective ALMA site are located
at a distance between 140\,pc and 200\,pc, such as Taurus, Lupus, Ophiuchus, Corona Australis, 
and Chamaeleon, we concentrate on simulations
at 900\,GHz in the following. 
Furthermore, even more distant star-forming regions (e.g., in Aquila, Vulpecula,
Vela, Perseus, Puppis, and Orion) are considered by comparison of simulated 900\,GHz observations
of the evolved disk seen under different inclinations in distances up to 300\,pc
in Fig.~\ref{pic1-all3}. While the main characteristics of the vortex (pre-planetary matter concentration), 
such as
its spatial extent and brightness contrast to the surrounding medium clearly can be extracted
from the images of objects at a distance of 140\,pc, even the spiral pattern in the circumstellar
disk (which is another, weaker feature of the disk in this evolutionary state) could be
resolved in the case of a very close young stellar object such as TW~Hya.
At a distance of 300\,pc the vortex is visible at low inclinations of the disk only, while
it causes a slight asymmetry of the image at higher inclinations.

\begin{figure}[ht]
  \epsscale{0.55}
  \plotone{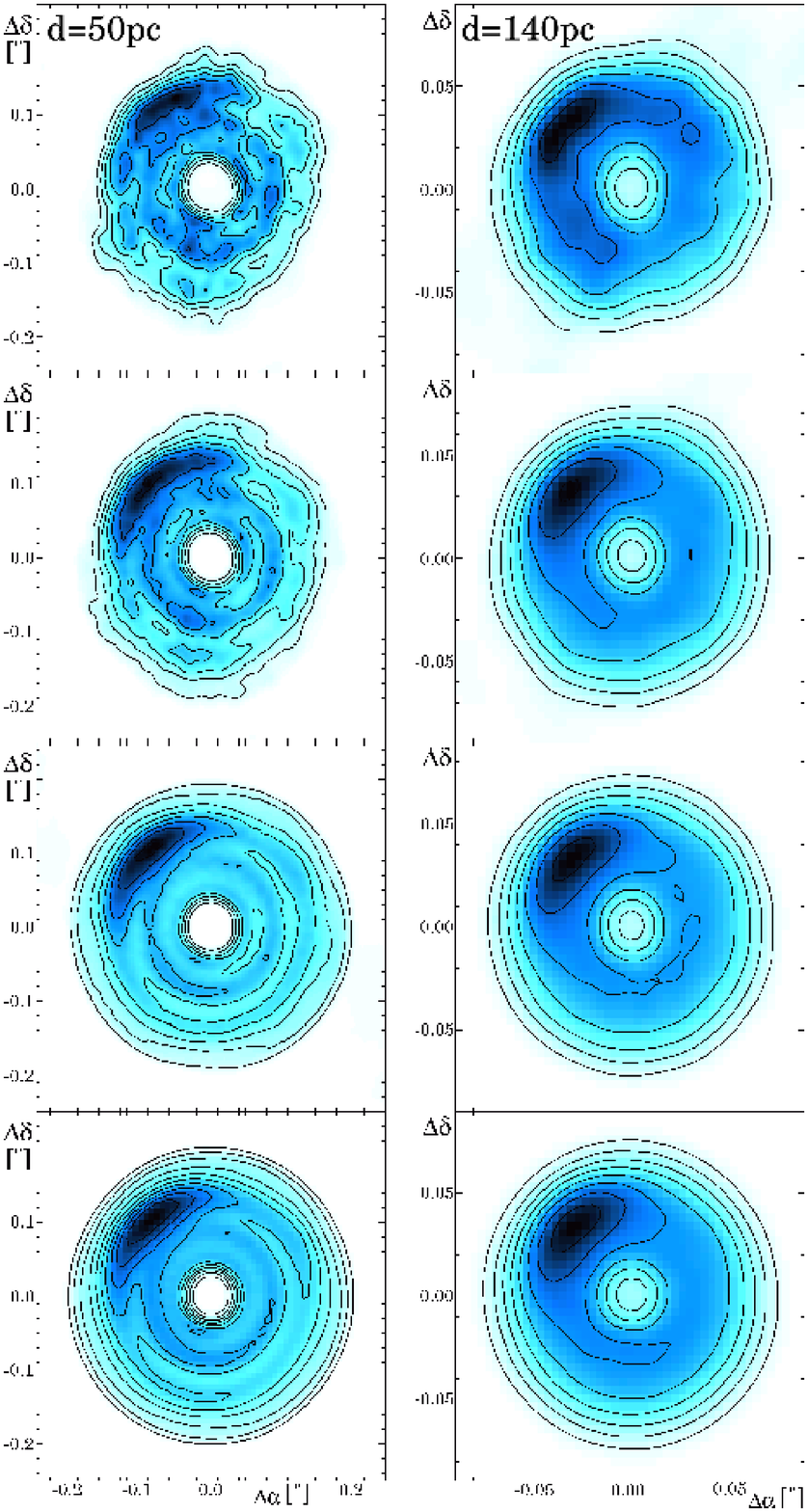}
  \caption{Reconstructed images of the evolved disk seen face-on in distances of 50\,pc and 140\,pc
    ($\nu$=900\,GHz).
    Total integration times of 5\,min, 24\,min, 2\,h and 8\,h are assumed
    (from top to bottom). The contour lines mark steps of 2\,mJy/beam.}
  \label{pic23-all3}
\end{figure}
Finally, we consider possible observing strategies (selected objects vs.\ surveys) 
which mainly results in the question of required total integration time per object.
In Fig.~\ref{pic23-all3} images of the evolved disk seen face-on resulting
from total integration times of 5\,min to 8\,h are shown.
It turns out that exposure times of 5\,min are sufficient to detect the vortex
and to derive its roughly parameters. In order to map the spiral density structures
in the case of an object at a distance of 50\,pc, an integration time of 1-2\,h is
required. A longer exposure time does not significantly increase the information content
of the reconstructed image. For objects in nearby star-forming regions (distance: 140\,pc),
a quick ``snapshot'' survey is sufficient since even much longer integration times will
not allow derivation of further details of the turbulent structure of the disk.

%-------------------------------------------------------------------------------
\section{Conclusions} \label{concl}

We found that ALMA will be able to map the anti-cyclonic hyper-dens vortices from the
 global baroclinic instability
representing an early stage of planet formation. In nearby star-forming regions,
the main parameters of the predicted global baroclinic instability in proto-planetary
disks can be derived within several minutes of integration time.
For even closer objects (within a distance of about 50\,pc) a detailed structure
of the spiral density pattern will be possible.
We want to stress that the largest planned baselines and high frequencies are
required for these observations.

%-------------------------------------------------------------------------------
\acknowledgments

We wish to thank F. Gueth for many helpful suggestions concerning the ALMA simulator
and L.\ Hillenbrand for careful reading of the manuscript.
S. Wolf was supported through the HST Grant GO\,9160, 
and through the NASA grant NAG5-11645. H.\ Klahr was supported by the NASA grant NAG5-4610,
the NSF grant AST 9987417, and by a special NASA astrophysics theory program which supports
a joint Center for Star Formation Studies at NASA-Ames Research Center, 
UC Berkeley, and UC Santa Cruz.

%-------------------------------------------------------------------------------

%-------------------------------------------------------------------------------
\end{document}